\documentclass
{mn2e}
\usepackage{epsfig}
\usepackage{amsmath, amssymb,bm}

\def \msun{\rm M_{\odot}}

\begin{document}
\title[How Big Can a Black Hole Grow?]{How Big Can a Black Hole Grow?}

\author[Andrew King ] 
{
\parbox{5in}{Andrew King$^{1, 2}$}
\vspace{0.1in} \\ $^1$ Department of Physics \& Astronomy, University
of Leicester, Leicester LE1 7RH UK\\ 
$^2$ Astronomical Institute Anton Pannekoek, University of Amsterdam, Science Park 904, NL-1098 XH Amsterdam, The Netherlands }

\maketitle

\begin{abstract}
I show that there is a physical limit to the mass of a black hole, above which it cannot grow through luminous accretion of gas, and so cannot appear as a quasar or active galactic nucleus. 
The limit is $M_{\rm max}\simeq 5\times 10^{10}\msun$ for typical parameters, but can reach 
$M_{\rm max}\simeq 2.7\times 10^{11}\msun$ in extreme cases (e.g. maximal prograde spin).
The largest black hole masses so far found are close to but below the limit. The Eddington luminosity $\simeq 6.5\times 10^{48}\,{\rm erg\, s^{-1}}$ corresponding to $M_{\rm max}$ is remarkably close to the largest AGN bolometric luminosity so far observed. The mass and luminosity limits both rely on a reasonable but currently untestable hypothesis about AGN disc formation, so future observations of extreme SMBH masses can therefore probe fundamental disc physics. Black holes can in principle grow their masses above $M_{\rm max}$ by non--luminous means such as mergers with other holes, but cannot become luminous accretors again. They might nevertheless be detectable in other ways, for example through gravitational lensing.
I show further that black holes with masses $\sim M_{\rm max}$ can probably grow above the values specified by the black--hole -- host--galaxy scaling relations, in agreement with observation.

\end{abstract}


\footnotetext[1]{E-mail: ark@astro.le.ac.uk}

\section{Introduction}
\label{intro}
Astronomers generally agree that the centre of almost every galaxy contains a supermassive black hole, with masses $M \sim 10^5 - 10^{10}\msun$. The observed hole masses correlate tightly with large--scale properties of the host galaxy's central bulge (see Kormendy \& Ho 2013 for a recent review). This initially surprising connection arises because the gravitational potential energy released as a black hole grows offers 
the
most efficient way of extracting energy from ordinary matter (e.g. Frank et al., 2002), and could potentially destabilize a host galaxy's central bulge (King, 2003). The huge luminosities they produce make accreting black holes detectable as quasars and active galactic nuclei (AGN), and exert mechanical feedback on their surroundings. This regulates their growth rates, and limits their masses to values specified by properties of the host (for a recent review see King \& Pounds, 2015).  As the host galaxies grow their masses, the black holes can grow further. An obvious question is whether there is a limit to this process, or whether a black hole can in principle reach any given mass, given a suitable host and enough time. The first attempt to answer this question was made by Natarajan \& Treister (2009). They derived a limit
by arguing that a selfÑgravitating accretion disc would blow itself away above a certain black hole mass of order $10^{10}\msun$, the precise value depending on the properties of the host galaxy's dark matter halo, in agreement with two lines of observational evidence.

I study this question further here. In Section 2 I give a simple argument for a physical limit on any SMBH mass,above which it cannot form a disc and so grow by luminous accretion, together with some direct consequences. In Section 3 I consider the effects of changing some of the assumptions made in the  simple argument of Section 2, and Section 4 is a discussion.
\begin{figure*}
\centerline{\psfig{file=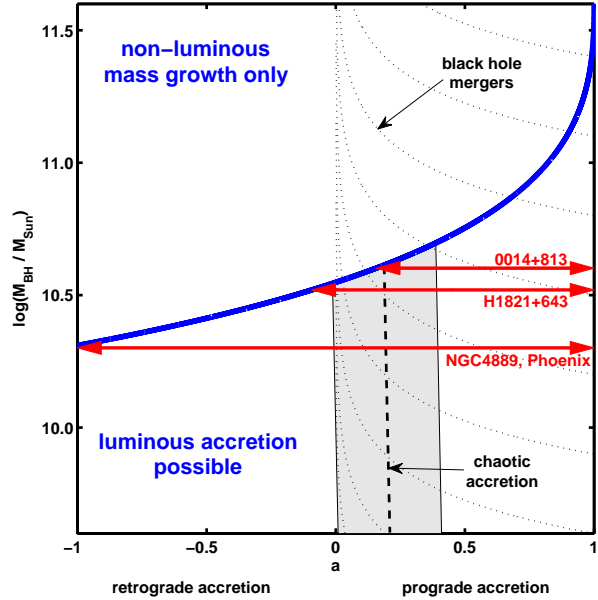,width=0.5\textwidth,angle=0}}
\caption{The mass limit $M_{\rm max}$ for accreting supermassive black holes, compared with the largest observed masses. The curve shows $M_{\rm max}$ as a function of black hole spin parameter $a$, where values $a < 0$ denote retrograde accretion. 
Accreting SMBH must lie below the curve. In order of decreasing mass, the systems shown are $0014+813$, with $M = 4\times 10^{10}\msun$ (Ghisellini et al., 2010),
the central quasar of the H$1821+643$ cluster ($M = 3\times 10^{10}\msun$, Walker et al., 2014), NGC 4889 (McConnell et al., 2011), and the central galaxy of the Phoenix cluster (McDonald et al., 2012), both of these systems having $M = 2\times 10^{10}\msun$. Only the first two systems place any restrictions on the black hole spin. For  $0014+813$, accretion must be prograde, with $a \gtrsim 0.2$. For H$1821+643$,
if accretion is prograde $|a|$ can have any value, but if accretion is retrograde $|a|$ must be less than about 0.1. All the other systems are compatible with any spin value. 
The dotted curves show the statistical effect $|a| \propto M^{-2.4}$ (Hughes \& Blandford, 2003) of black hole mergers on mass growth. The attractor $|a| \longrightarrow \bar a \simeq 0.20M_{10}^{-0.048}$ 
for chaotic gas accretion is shown (dotted track) with the grey surround indicating the typical spread in $|a|$ (King, Pringle \& Hofmann, 2008).
}
\label{}
\end{figure*}

\section{Supermassive Black Hole Growth}
\label{smbh}
\subsection{Disc Accretion in AGN}
It has long been known that supermassive black holes (SMBH) grow their masses mainly by luminous accretion of gas (Soltan, 1982). Since gas within galaxies must have significant angular momentum, SMBH accretion must at any given time proceed largely through a disc (more probably, a series of discs of varying orientation). 
Any SMBH disc is likely to be self-gravitating outside
some radius $R_{\rm sg} \sim 0.01 - 0.1$\,pc (Shlosman et al. 1990; Collin--Souffrin \& Dumont 1990; Hur\'e et al. 1994). To see this I note that
the condition for stability against self--gravity for a gaseous disc can be expressed as  
\begin{equation}
\frac{c_s\Omega}{\pi G\Sigma} >1
\label{toomre}
\end{equation}
(Toomre, 1964), where $c_s, \Omega$ and $\Sigma$ are the local sound speed, orbital frequency and surface density respectively, and $G$ is the gravitational constant. Vertical force balance in a disc requires $c_s = H\Omega$, where $H$ is the disc semithickness (e.g. Pringle, 1981; Frank et al., 2002). Using this in (\ref{toomre}) gives the stability requirement
\begin{equation}
\rho < \frac{\Omega^2}{2\pi G} = \frac{M}{2\pi R^3}.
\label{rhosg}
\end{equation}
Here  $\rho = \Sigma/2H$ is the mean density of the disc, and I have used the Keplerian relation $\Omega = (GM/R^3)^{1/2}$, with $R$ the local disc radius, as appropriate for a thin disc around a black hole of mass $M$ (Pringle, 1981; Frank et al., 2002) at the last step.
The disc mass interior to radius $R$ is
$M_d \simeq 2\pi R^2H\rho$, so (\ref{rhosg}) can be expressed in the well--known form 
\begin{equation}
M_d\lesssim \frac{H}{R}M
\label{msg}
\end{equation}
(Pringle, 1981). 

For the parameters appropriate to thin discs around SMBH (I discuss this in more detail in Section~3 below) the disc aspect ratio obeys
$H/R \sim 10^{-3}$ (cf Collin--Souffrin \& Dumont, 1990; King, Pringle \& Hofmann 2008). The full disc equations then give
\begin{equation}
R_{\rm sg} = 3\times 10^{16}\alpha_{0.1}^{14/27}\eta_{0.1}^{8/27}
(L/L_{\rm Edd})^{-8/27}M_8^{1/27}\, {\rm cm}
\label{rsg}
\end{equation}
where $L$ is the accretion luminosity and $L_{\rm Edd}$ the Eddington luminosity
(cf Collin--Souffrin \& Dumont, 1990; King \& Pringle 2007).
Here $\alpha = 0.1\alpha_{0.1}, \eta = 0.1\eta_{0.1},  \dot m = \dot M/\dot M_{\rm Edd}$ are the standard viscosity parameter, the accretion efficiency and Eddington accretion ratio respectively, and $M_8 = M/10^8\msun$.
Gas cooling in the outer regions of these discs is fast enough that
self--gravity is likely to lead to star formation rather than increased
angular momentum transport (Shlosman \& Begelman 1989; Collin
\& Zahn 1999). The very small aspect ratios $H/R$ and
masses $M_d$ of AGN discs mean 
that self--gravity appears first in modes with azimuthal wavenumber
$m \simeq R/H \sim 10^3$. These produce transient spiral waves which initially transport angular momentum (Anthony \& Carlberg 1988; Lodato \& Rice 2004,
2005). In a disc which is locally
gravitationally unstable in this way, most
of the gas initially at radii $R > R_{\rm sg}$ is likely either to form into stars, or
to be expelled by those stars which do form, on a near--dynamical
timescale (cf Shlosman \& Begelman 1989).

The estimate (\ref{rsg}) is almost independent of parameters, and so must apply to almost every SMBH. Encouragingly, $R_{\rm sg}$  is only slightly smaller than the inner edge  
$\sim 0.03$~pc of the ring of young stars seen around the black hole in the centre of the Milky Way (Genzel et al., 2003), strongly suggesting that the most recent accretion event on to the central SMBH formed an accretion disc within  $R_{\rm sg}$ and passed its angular momentum to the self-gravitating region further out which produced these stars. The small disc masses $M_d \la 10^{-3}M$ expected from the self--gravity constraint offer an immediate explanation of the preferred timescale $\sim 10^5$~yr of SMBH growth phases (Schawinski et al. 2015; King \& Nixon, 2015).

This reasoning implies that the outer radius of any SMBH accretion disc cannot exceed $R_{\rm sg}$, which is effectively independent of the SMBH mass. But the inner disc radius must be at least as large as the ISCO (innermost stable circular orbit) around the SMBH, whose size scales directly with the SMBH mass M, as
\begin{equation}
R_{\rm ISCO} = f(a)\frac{GM}{c^2} =  7.7\times 10^{13}M_8f_5\,{\rm cm}.                
\label{isco}
\end{equation}  
Here $f(a)$  is a dimensionless function of the SMBH spin parameter $a$, with
$f(a) = 5f_5(a)$, so that $f_5 \simeq 1$ corresponds to prograde accretion at moderate SMBH spin rates $a \simeq 0.6$. If $R_{\rm ISCO}\gtrsim R_{\rm sg}$, disc accretion is likely to be suppressed. Any disc material arriving at $R_{\rm ISCO}$ feels only very weak outward angular momentum transport. If the SMBH accretes any of this matter at all, it must be self--gravitating and so swallowed whole, without radiating as a disc. An SMBH might in principle grow its mass in this way, but we shall see below that it cannot subseqently reappear as a bright disc--accreting object, i.e. a quasar or AGN.

\subsection{SMBH Mass Limit}
Comparing (\ref{rsg}) and (\ref{isco}) we see that the ISCO radius exceeds the self--gravity radius, making disc formation impossible, for SMBH masses larger than
\begin{equation}
M_{\rm max} = 5\times 10^{10}\msun\alpha_{0.1}^{7/13}\eta_{0.1}^{4/13}
({\it L/L}_{\rm Edd})^{-4/13}{\it f}_5^{-27/26}
\label{mmax}
\end{equation}  
This is a physical upper limit to the mass of the SMBH in any quasar or AGN, since these systems have accretion discs. 

Figure 1 shows the curve $M = M_{\rm max}(a)$, with $\alpha = 0.1$ and
$L = L_{\rm Edd}$, while $\eta, f_5(a)$ are specified parametrically as functions of $a$ (cf the relations 9 and 11 in King \& Pringle, 2006). The whole curve lies  
slightly above all the masses measured for accreting SMBH except for $0014+813$
(Ghisellini et al., 2010: $M \simeq 4\times 10^{10}\msun$) and
H1821+643 (Walker et al., 2014: $M \simeq 3\times 10^{10}\msun$). The first system is compatible with the limit provided that accretion is prograde and $a\gtrsim 0.2$. H1821+643
is compatible with the limiting mass provided that $a \gtrsim - 0.1$ -- that is, prograde accretion is possible for any spin parameter $a >0$, but retrograde accretion on to this hole with $|a| > 0.1$ is ruled out.

For spin rates $a = 1$  corresponding to maximal prograde spin wrt the sense of accretion, the normalization in (\ref{mmax}) becomes $2.7\times 10^{11}\msun$, which is the absolute maximum for an accreting SMBH. In practice values of $M$ of this order are likely to be rather rare, as this requires disc accretion to be almost permanently prograde as the hole mass grows (the spin--down effect of retrograde accretion is greater than spin--up by prograde accretion, because of its larger ISCO and so its lever arm). This in turn probably requires the hole spin to be permanently correlated with a fixed direction of the potential controlling gas flow within the galaxy, and so would tend to produce a spin axis and hence AGN jet direction which is similarly aligned with the galaxy. Observations do not support this predicted correlation (Nagar \& Wilson, 1999; Kinney et al., 2000, Sajina et al., 2007). 

If accretion is not controlled by a large--scale potential in this way, it presumably has to involve multiple small--scale events, essentially random in time and orientation. This `chaotic' type of accretion (King \& Pringle 2006, 2007; King \& Nixon, 2015)
%
%
leads statistically to spin--down, again because retrograde events have larger lever arms than prograde, and occur almost as often. King, Pringle \& Hofmann (2008) show that this type of feeding predicts an attractor $|a| \longrightarrow \bar a \simeq 0.20M_{10}^{-0.048}$ for large SMBH masses (shown in Fig. 1). Other interactions with the SMBH also tend to reduce $|a|$. In particular, mergers with other black holes statistically decrease the spin as $|a| \propto M^{-2.4}$ (Hughes \& Blandford, 2003). All these considerations  suggest that SMBH usually cross the critical $M = M_{\rm max}$ curve (\ref{mmax}) at modest values of $|a|$, so that $M_{\rm max} \simeq 5 \times 10^{10}\msun$ in all but rare cases.

\subsection{AGN Luminosity Limit}
Assuming that the maximum observable luminosity of an AGN obeys the Eddington limit, the mass limit (\ref{mmax}) implies a luminosity limit
\begin{equation}
L_{\rm max} = 6.5 \times 10^{48}\alpha_{0.1}^{7/13}\eta_{0.1}^{4/13}
{\it f}_5^{-27/26}\,{\rm erg\,s^{-1}}
\label{lmax}
\end{equation}  
This prediction is in remarkably good agreement with the highest QSO bolometric luminosity
found in the recent WISE survey of hot, dust--obscured galaxies (`Hot DOGs': Assef et al., 2015, Figure 4), which appears to be the largest AGN luminosity so far found. (The SEDs of Hot DOGs are generally dominated by a luminous obscured AGN.)

\section{Raising the Limit?}

The mass and luminosity limits (\ref{mmax}, \ref{lmax}) agree well with current observations. Their derivations given above are simple enough that it is worth asking about the effect of varying some of the assumptions made there.

First, we should note that (\ref{mmax}, \ref{lmax}) are the {\it observable} limits for luminous accretion at the luminosity $L$, not absolute limits on the black hole mass (as I have remarked already, {\it non}--luminous mass growth beyond $M_{\rm max}$ is perfectly possible).
{\it Sub}--luminous mass growth (i.e. at lower accretion rates, with $L/L_{\rm Edd} < 1$) beyond $M_{\rm max}$ is also possible. Such systems would by definition be fainter and so harder to detect. A flare taking the luminosity briefly up to $L_{\rm Edd}$ does not help, as the SMBH mass would then exceed the $M_{\rm max}$ value corresponding to this brighter state and so not be detectable (physically, the higher accretion rate moves the self--gravity radius inside the ISCO, preventing luminous accretion). In a very similar way, other subluminous accretion, e.g. via an ADAF, does not get around the observable limit.

The most radical way around the limits (\ref{mmax}, \ref{lmax}) is fairly obvious on looking at eqn (\ref{msg}). Luminous disc accretion with a significantly larger scaleheight $H$ than the very small values ($H \sim 10^{-3}R$) for standard thin--disc AGN accretion could increase $M_{\rm max}$ dramatically. The most likely way this could potentially occur is if radiation pressure would dominate gas pressure in the outer parts of the disc, i.e. if much of the outer disc would lie in the region `a' of Shakura and Sunyaev, rather than region `b' (gas pressure dominant) as assumed above. The discussion by Kawaguchi et al. (2004) shows that this might happen near the self--gravity radius if the prevailing accretion rate satisfies
\begin{equation}
\dot M > \dot M_{\rm ba} = 4\alpha_{0.1}^{0.4}\,\msun\,{\rm yr^{-1}}
\label{ba}
\end{equation}
which is sub--Eddington for SMBH masses $M \ga 4\times 10^8\alpha_{0.1}^{0.4}\msun$. But is is well known that disc region `a' is strongly unstable on a thermal timescale, both in the context of the $\alpha$--prescription (Lightman \& Eardley, 1974) and in shearing--box simulations (Jiang et al., 2013). 
Theory and simulations are not currently able to work out the consequences of this instability, so
we should ask what observational constraints exist. If region `a' is fed material from a disc region `b' outside it, it evidently finds a way to supply matter to the black hole. We observe Eddington--limited systems at a wide range of black hole masses, supplied by a stable and longlasting reservoir such as a companion star in a stellar--mass binary, so an inner region `a' is perfectly compatible
with AGN feeding. But it is much less obvious that an SMBH disc can {\it form} with region `a' conditions at its outer, self--gravitating radius, and stably feed the hole. An AGN feeding event probably involves a ballistic flyby of a mass of gas, dust and possibly stars, which becomes bound to the SMBH because of internal dissipation as tides act on it (cf King \& Pringle 2006, King \& Nixon 2015). With radiation pressure already dominant at the self--gravity radius this dissipation seems likely to drive mass off rather than produce efficient AGN feeding. If accordingly we assume that the outer parts of discs feeding AGN cannot be in region `a' conditions we are again left with the limits (\ref{mmax}, \ref{lmax}).

\section{Discussion}

There are several points to note about the limits $M_{\rm max}, L_{\rm max}$.

1.  Once $M > M_{\rm max}$ the SMBH can still go on growing its mass, as I have remarked above, provided this does not involve luminous disc accretion. Indeed gas accretion by swallowing stars would produce very little radiation, removing the Eddington limit as a barrier to growth. But the arguments above suggest that this mass growth is unlikely to allow the black hole to reappear as an accreting quasar. An increase in $|a|$ is unlikely, and disc accretion is in any case impossible for $M > 2.7\times 10^{11}\msun$ for any value of 
$a$. One might nevertheless detect SMBH with masses above $M_{\rm max}$ in other ways, perhaps through gravitational lensing for example.

2.  At masses close to but below $M_{\rm max}$, luminous disc accretion in a field galaxy is likely to approach the Eddington luminosity only rarely, since even the dynamical infall rate $f_g\sigma^3/G$  (with $f_g$ the gas fraction) is below the Eddington value except in galaxy bulges with very high velocity dispersions $\sigma \gtrsim 400\,{\rm km\, s^{-1}}$. Even for brightest cluster galaxies (BCGs) in the centres of clusters, accretion of cluster gas may be vigorous, but strongly super--Eddington rates appear unlikely.
The sub--Eddington accretion likely to prevail in such systems does not trigger the strong feedback which probably underlies the $M - \sigma$ relation, so SMBH close to $M_{\rm max}$ can evolve above $M - \sigma$, and need not make their hosts red and dead. The observational data (e.g. McConnell et al., 2011) suggest that this is indeed what happens.

3.  In line with this,  $M_{\rm max}$ lies well above the $M - \sigma$  relation, since host bulge velocity dispersions do not reach the required values $\sigma \gtrsim 700\,{\rm km\, s^{-1}}$. 

4. The limits $M_{\rm max}, L_{\rm max}$ might be breached if AGN discs could form with radiation pressure dominant at the self--gravity radius. So the survival of this limit in the face of further observations may have something to tell us about AGN disc formation, and indeed the nonlinear development of the radiation pressure instability in these discs.

\section*{Acknowledgments}

I thank Priya Natarajan for comments, 
Chris Nixon and Stuart Muldrew for discussions and help in preparing the paper, and the referee for a very perceptive report. This paper was completed during a visit to the Institut d'Astrophysique, Paris, and I am grateful to them for support. I particularly thank Jean--Pierre 
Lasota for discussions of accretion disc stability during my stay. Theoretical astrophysics research in Leicester is supported by an STFC Consolidated Grant.

{}

\end{document}